# Control of the chemiluminescence spectrum with porous Bragg mirrors

Simone Varo[1,2]§, Luigino Criante[2]§, Luca Passoni[1,2], Andrea Delle Vedove[2,3], Eduardo Aluicio Sarduy[1,2], Fabio Di Fonzo[2], Guglielmo Lanzani[1,2]*, Francesco Scotognella[1,2,4]*

[1] Dipartimento di Fisica, Politecnico di Milano, Piazza L. da Vinci 32, 20133 Milano, Italy
[2] Center for Nano Science and Technology@PoliMi, Istituto Italiano di Tecnologia, Via Giovanni Pascoli, 70/3, 20133 Milano, Italy
[3] Dipartimento di Chimica, Materiali e Ingegneria Chimica "Giulio Natta", Politecnico di Milano, Via Mancinelli 7, 20131 Milano, Italy
[4] Istituto di Fotonica e Nanotecnologie CNR, Piazza L. da Vinci 32, 20133 Milano, Italy

**Abstract**
Tunable, battery free light emission is demonstrated in a solid state device that is compatible with lab on a chip technology and easily fabricated via solution processing techniques. A porous one dimensional (1D) photonic crystal (also called Bragg stack or mirror) is infiltrated by chemiluminescence rubrene-based reagents. The Bragg mirror has been designed to have the photonic band gap overlapping with the emission spectrum of rubrene. The chemiluminescence reaction occurs in the intrapores of the photonic crystal and the emission spectrum of the dye is modulated according to the photonic band gap position. This is a compact, powerless emitting source that can be exploited in disposable photonic chip for sensing and point of care applications.

**Introduction**
The emission of electromagnetic radiation via release of energy in a chemical reaction is called chemiluminescence (CL) [1,2]. The steps of the light generating process are shown in Scheme 1 [3].

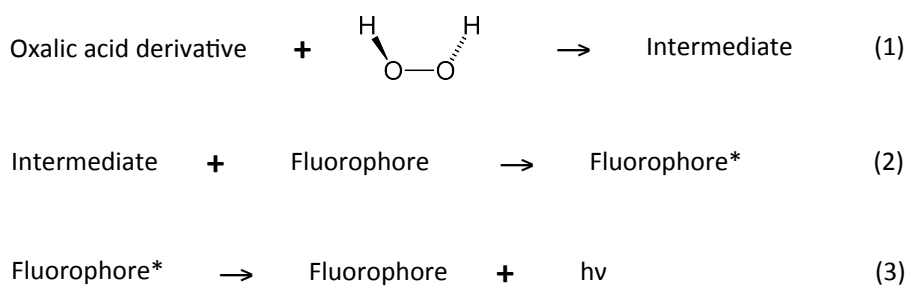

**Scheme 1.** Example of the chemiluminescence process.

In reaction (1) an oxalic acid derivative reacts with hydrogen peroxide, resulting in an intermediate highly energetic molecular species. In reaction (2) the intermediate transfers its energy to an acceptor fluorophore, which becomes electronically excited and emits its energy in form of light, as sketched in scheme 1(3).
In opto-chips, where light signals are generated by integrated sources such as LEDs and lasers, a substantial fraction of the energy consumed is employed for powering the emitter. The exploitation of CL can lead to a consistent reduction of the electrical energy consumption, since CL is self-sustained and it does not require an external power source. In general a battery free light emitter can be used in a manifold of applications such as lab on a chip and other integrated photonic circuits, highly sensitive portable sensors, new concept displays and integrated communication boards. On the other hand, the main drawback of CL are the problematic confinement in space of the reaction and its low quantum efficiency. For this reasons, many research groups are developing novel CL methods, acting on the reaction conditions, changing reactants or adding catalysts [4–9]. A better way to approach the problem would however be to exploit different mechanisms of excitation and process control, not only from a chemical point of view. An example still to be consolidated is the exploitation of plasmonic effects [10] while the use of optical confinement in photonic crystals seems the most promising [11,12]. In the last years, porous one-dimensional photonic crystals (stacks of alternated porous layers, also called Bragg mirrors [13–20]) have been a very active research topic in photonics. They provide a very efficient photonic band gap (i.e. with the transmission that tends to zero) that is predictable in analytical ways, as the transfer matrix method [21]. Porosity is an additional feature that can be obtained by using proper materials and fabrication processes [22,23]. These materials can be good candidates for low-cost optical sensing [24–27]. Furthermore, when the porous materials are infiltrated with different types of active materials, it is possible to realize organic light emitting diodes with enhanced luminescence [28], organic lasers [29–31], electro-optic switches [32,33] and dye sensitized solar cells with enhanced light absorption [34–36].

In this work, we demonstrate CL emission modulation by the employment of nanoparticle based porous one dimensional Bragg mirrors. Such one-dimensional photonic crystals have been infiltrated with the standard reagents of the CL solution. The CL occurs in the photonic crystal and the emission spectrum of the dye is modulated by the photonic band gap of the structure. In this way one can combine and separately tune the optical properties of the structure (via its fabrication parameters) and of the chemical process (via the choice of the reactants and of the fluorophore). Encounter, intermediate formation and radiative decay take place in the confined micro environment defined by the cavities in the porous medium: this approach is thus highly preferable for integration in solid state devices as photonic crystals technology is well established.

We have employed two different types of fabrication for the Bragg mirrors, i.e. spin coating and pulse laser deposition, to demonstrate the possibility of an easy integration of a chemiluminescent process in a layered photonic structure. Our application is thus a substantial upgrade of an existing photonic component that allows for a reduction in the energy consumption of the device in which it is integrated, without losing other established features. This is a step towards a powerless lab on a chip.

**Experimental details**

The two CL reagent solutions used in the experiment were prepared accordingly to the general procedure described in the original patent [37]. The first solution was made by mixing in a volumetric glass flask 9 mg of the rubrene fluorophore and 90 mg of bis(2,4,6 trichlorophenyl)oxalate, then filling with ethyl acetate to a volume of 20 ml. The second solution was made by filling a volumetric glass flask containing 480 mg of sodium salicylate and 10 ml of 30% hydrogen peroxide solution to a volume of 20 ml using methanol. All components were purchased from Sigma-Aldrich.

In preparation for our experiment, a 7 bilayers (i.e. 14 layers) Bragg mirror was fabricated by means of a spin coating technique. The starting materials employed being a colloidal solution of silica particles with an average size of 10-15 nm (Sigma Aldrich Ludox SM-30), further diluted in water to a concentration of 5% in weight, and a solution of titanium dioxide nanoparticles with an average size of 20 nm, obtained by diluting in ethanol a commercially available paste (Dyesol 18 nr-t) to a 1:9 weight ratio. The crystal was therefore made by alternated deposition of the solutions on a glass substrate, previously cleaned in acetone and isopropanol sonic baths lasting approximately 10 minutes each. Every deposition was followed by a spinning of the substrate at a speed of 3800 RPM for 60 seconds using a Laurell WS-400-6NPP-Lite spin coater, and by drying in air at a temperature of $350^0$ C for 20 minutes. Moreover, two 8 bilayer (i.e. 16 layers) Bragg mirrors were fabricated by pulse laser deposition, which allows to obtain different layer (in our case of titanium dioxide) with different porosity, resulting in a significant refractive index contrast. The detailed experimental procedure has been published elsewhere [38].

The optical characterization of the Bragg stacks has been performed by using a tungsten lamp and concave grating spectrometer, acquired from Stellarnet (spectral resolution 1.5 nm). We used a linearly polarized light for the optical characterization of the photonic structure (even if we have observed that, in our experiments, light polarization has not significantly affected the optical properties of the structure).

**Results and Discussion**

A schematic of the CL reaction in the porous photonic crystal is shown in Figure 1. With two syringes we have infiltrated the peroxide solution and the fluorophore-containing one, prepared as described previously, in the pores formed between the nanoparticles composing the layers of the photonic crystal. The two solutions, after a percolation along the pore network, coalesce resulting in a bright CL.

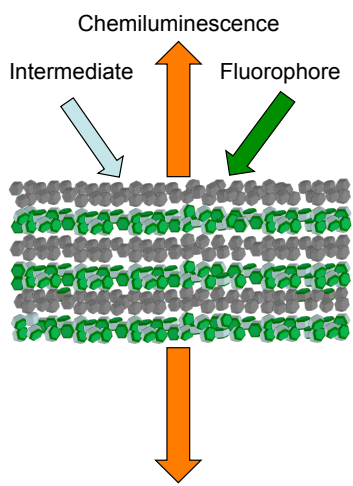

**Figure 1.** Schematic of the CL reaction in the porous one-dimensional photonic crystal. The CL reagents are infiltrated at the same time in the porous structure. The CL spectrum will be modulated by the photonic structure.

The fabricated photonic crystals show band gaps at about 550 nm. The spectral position of the photonic band gap is due to the thicknesses of the layers and their effective refractive indexes. The effective refractive index in the bare photonic crystal is related to the refractive indexes of metal oxide and air, while in the infiltrated photonic crystal the effective refractive index is influenced by the filling of the pores with the CL reagent solutions. For this reason, the infiltrated photonic crystals show a red shift of 40-45 nm of the photonic band gap with respect to the one of the bare photonic crystal (Figure 2).

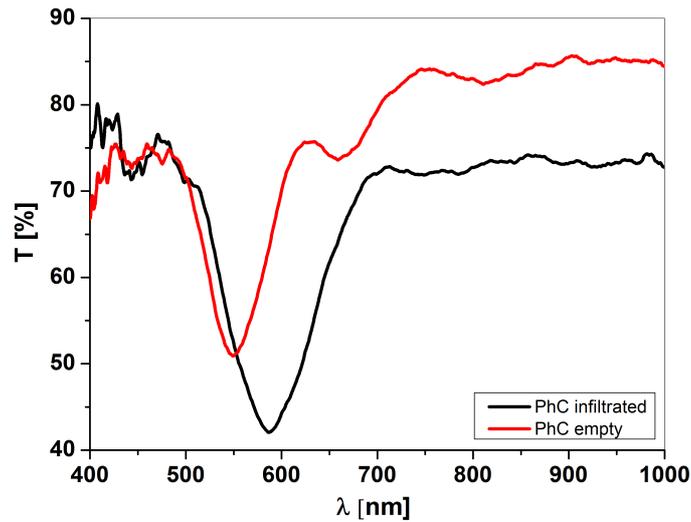

**Figure 2.** Transmission spectrum of the porous one-dimensional photonic crystal before and after infiltration of the CL reagents. We observe a shift of 40-45 nm with infiltration, due to the change of the effective refractive index of the photonic crystal layers.

The CL in different photonic crystals is shown in Figure 3. The photonic crystals display different photonic band gaps, resulting in a diverse modulation of the fluorophore (in this case rubrene) emission. In Figure 3a the CL is modulated by a nanoparticle-based photonic crystal, while in Figure 3b and 3c the CL is modulated by two porous photonic crystals with alternated layers of different porosity, made by pulse laser deposition.

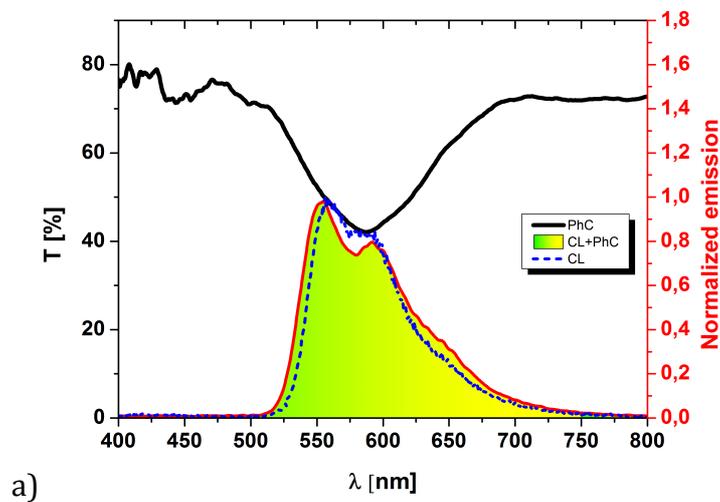

a)

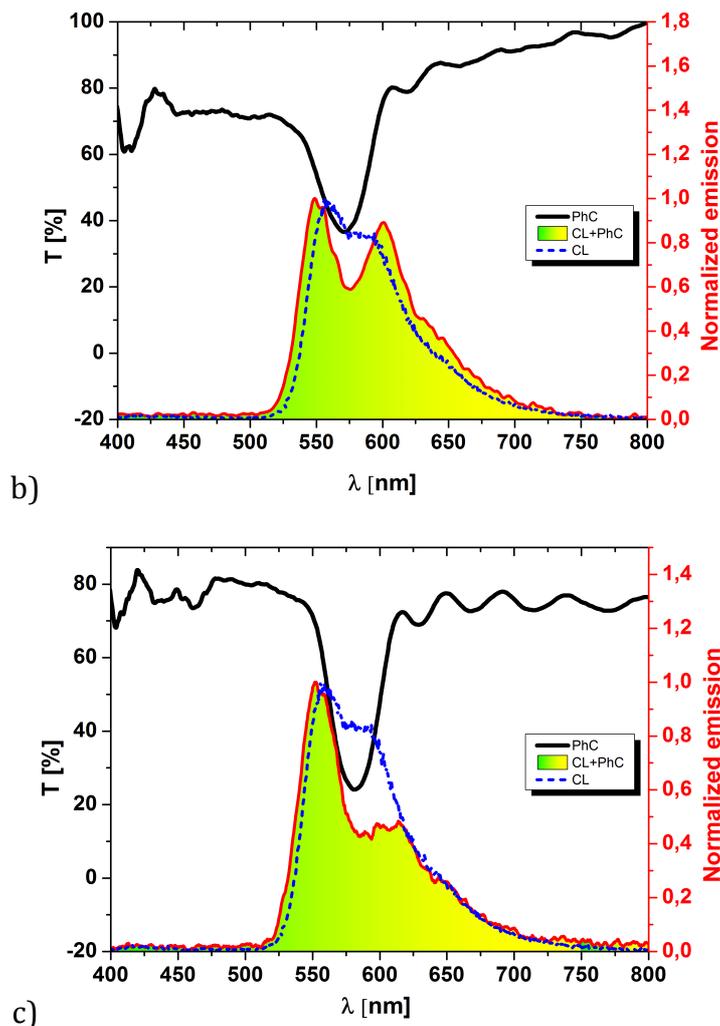

**Figure 3.** CL emission with rubrene filtrated by the one-dimensional porous photonic crystal made by a) spin coating technique and b) and c) pulse laser deposition with two different porosities of the layers. The black curves show the transmission spectra of the photonic crystals. The blue dashed curves depict the CL spectrum of rubrene, while the red curves represent the CL spectra of rubrene with the CL reagents infiltrated in the porous photonic crystals.

The photonic crystals, if properly designed, and with a high quality factor, can provide a feedback to the CL intensity, resulting in a subsequent emission enhancement, as was already demonstrated for opals [39].

**Conclusions**

In this paper we have demonstrated CL emission in a porous one dimensional Bragg mirror. This was achieved by infiltrating the photonic structure with the CL reagents. The CL spectrum is modulated by the photonic band gap, providing a handle for spectral tuning of the emission. The proposed system provides efficient emission without an electrical bias or an external pumping light source. This paves the way for the fabrication of portable battery-free light sources, displays active pixels and microfluidic chips for sensing applications. A possible lab on a chip could employ the integration of microchannels that inject the chemiluminescence reagents in the porous Bragg mirrors.

**Author Information**

* To whom the correspondence should be addressed:
Francesco Scotognella, Dipartimento di Fisica, Politecnico di Milano, piazza Leonardo da Vinci 32, 20133 Milano, Italy. Tel.: +390223996056. Email address: francesco.scotognella@polimi.it
Guglielmo Lanzani, Center for Nano Science and Technology@PoliMi, Istituto Italiano di Tecnologia, Via Giovanni Pascoli, 70/3, 20133 Milano, Italy. Tel.: +390223999872. Email address: guglielmo.lanzani@iit.it
§ S. V. and L. C. equally contributed to this work.

**Acknowledgements**
F.S. acknowledges financial support from Italian Ministry of University and Research (project PRIN 2010-2011 "DSSCX", Contract 20104XET32).